\documentclass[12pt]{iopart}
\usepackage{rotating,graphics,graphicx,subfigure,amssymb,xcolor}

\begin{document}

\title[Contact value theorem for interfaces with modulated surface charge density]{Contact value theorem for electric double layers with modulated surface charge density}

\author{Ladislav \v{S}amaj}

\address{Institute of Physics, Slovak Academy of Sciences, 
D\'ubravsk\'a cesta 9, 84511 Bratislava, Slovakia}
\ead{Ladislav.Samaj@savba.sk}
\vspace{10pt}
\begin{indented}
\item[]
\end{indented}

\begin{abstract}
The contact value theorem was originally derived for Coulomb fluids
of mobile charged particles in thermal equilibrium, in the presence of
interfaces carrying a {\em uniform} surface charge density and in
the absence of dielectric discontinuities.
It relates the pressure (the effective force) between two parallel
electric double layers to the particle number density and
the surface charge density at the interface,
separately for each of the two electric double layers.
In this paper, we generalise the contact value theorem to electric double
layers with interfaces carrying a {\em modulated} surface charge density.
The derivation is based on balance of forces exerted on interfaces.
The relevance of particular terms of the contact value theorem is tested
on an exactly solvable two-dimensional Coulomb system with counterions only
at the coupling constant $\Gamma=2$.
\end{abstract}


\vspace{2pc}


\submitto{\jpa}

\maketitle

\renewcommand{\theequation}{1.\arabic{equation}}
\setcounter{equation}{0}

\section{Introduction} \label{Section1}
The study of the thermal equilibrium of classical (i.e., non-quantum) fluids
of charged particles interacting pairwisely via Coulomb potential
is important in soft and condensed matter physics \cite{Levin02}.

The Coulomb potential can be defined in any Euclidean space of points
${\bf r}=(x_1,x_2,\ldots,x_{\nu})$ (spatial dimensions $\nu=1,2,3,\ldots$).
In Gauss units and in vacuum with dielectric constant $\varepsilon=1$,
it corresponds to the solution of the $\nu$-dimensional Poisson equation:
\begin{equation}
\Delta v({\bf r}) = - s_{\nu}\delta({\bf r}) ,
\end{equation}
where $\Delta=\sum_{j=1}^{\nu} \partial^2/\partial x_j^2$ is the
$\nu$-dimensional Laplacian, $\delta$ is Dirac's delta function and 
$s_{\nu}=2\pi^{\nu/2}/\Gamma(\nu/2)$ the surface area of
the $\nu$-dimensional unit sphere. 
In particular,
\begin{equation} \label{Coulomb}
v({\bf r}) = \left\{
\begin{array}{ll}
-r & \mbox{for $\nu=1$} ,\cr 
\ln (r/L) & \mbox{for $\nu=2$} ,\cr 
r^{2-\nu}/(\nu-2) & \mbox{for $\nu\ge 3$} ,
\end{array} \right.  
\end{equation}
where $r$ is the modulus of ${\bf r}$ and $L$ is the free length scale.
In a three-dimensional (3D) space, the Coulomb potential has the
standard $1/r$ form known from electrostatics. 
The two-dimensional (2D) logarithmic potential can be interpreted in
real 3D space as the effective potential between parallel infinite
charged lines perpendicular to a plane, mimicking 3D polyelectrolytes.
Because the Fourier component of (\ref{Coulomb}) exhibits singular
behavior of type $\hat{v}({\bf k})\propto 1/k^2$, many generic properties 
of Coulomb fluids in thermal equilibrium like perfect screening are preserved
in any dimension $\nu$ \cite{Martin88}, leading to the well known
zeroth-moment and second-moment Stillinger-Lovett conditions for
the pair charge-charge densities \cite{Stillinger68,Carnie81}
in the bulk regime and to many other sum rules for the semi-infinite
geometries.
In dynamical systems of charged colloids exhibiting
diffusiophoresis in flow \cite{Anderson89,Shin17} or dielectrophoresis
in time-varying electric fields \cite{Zhou13}, the induced charge clouds
carry multipoles and therefore only a limited number of sum rules persists,
see section IV of Ref. \cite{Martin88}.
In this paper, we restrict ourselves to charged particle systems in thermal
equilibrium.
It should be emphasized that the fluid sum rules do not apply to large Coulomb
couplings characterized by a crystalline phase.

In biological experiments with macromolecules (colloids) immersed
in water or similar polar solvents, the colloidal surface acquires
a fixed surface charge density through the dissociation of microions
(counterions) into the solvent \cite{Evans99}.
The solvent contains ions generically of both signs, the corresponding
more-component systems are referred to as ``with salt added.''  
One can reach experimentally the deionized (salt-free) solvent
\cite{Raspaud00,Palberg04,Brunned04}, the corresponding one-component
models are referred to as ``with counterions only'' (or salt-free limit).
Mobile counterions are usually taken as electrons with the negative
elementary charge $-e$ while colloids are large balls whose surfaces
contain thousands of positive charges $e$ fixed at their positions.
To simplify the theoretical treatment of models, the curved surface
of the colloid is substituted by a planar one and the modulated charge
density on the surface by a uniform one.
The finite size of colloids is often ignored by taking the interior of
colloids as a semi-infinite walls, with no dielectric jump between
the medium the charges are immersed in and the colloid.
The charged surface in thermal equilibrium with surrounding mobile charges
form a neutral entity which is known as the electric double layer (EDL)
\cite{Gulbrand84,Attard88,Attard02,Messina09}.
The effective interaction of two EDLs, mediated by mobile microions,
is the topic of particular interest in soft matter \cite{Hansen00}. 

We shall concentrate on two basic versions of Coulomb fluids which are of
special experimental and theoretical interest:
\begin{itemize}
\item  
The jellium is a one-component plasma (OCP) of pointlike mobile particles with
the same (say elementary) charge $-e$, immersed in a fixed neutralizing
background charge distributed uniformly in space.
The definition of the pressure is not unique for the jellium because
of the presence of the rigid volume background charge \cite{Choquard80}.
There exists a version of one-component plasmas with the neutralizing
charge distributed not in the bulk, but at the surfaces of the walls of
the domain the mobile charges are confined to.
This is just the model for colloids with counterions only discussed above.
In what follows, we shall restrict ourselves to this kind of systems
for which, in contrast to jellium models, the pressure is defined uniquely.
The only relevant thermodynamic parameter in 2D one-component models,
which are in thermal equilibrium at the inverse temperature
$\beta=1/(k_{\rm B}T)$, is the coupling constant
\begin{equation} \label{coupling}
\Gamma\equiv \beta e^2 .
\end{equation}
\item
The symmetric two-component plasma (TCP), or Coulomb gas, is the overall
neutral system of $\pm e$ charges with a hard core which prevents from
a thermodynamic collapse of opposite charges.
The hard core is not necessary in the 2D Coulomb gas when the Boltzmann
factor corresponding to the interaction of opposite $\pm e$ charges,
$r^{-\Gamma}$, is integrable at small 2D distances, i.e. for $\Gamma<2$.
The TCP in the presence of charged walls corresponds to colloids with salt
added in solvent discussed above.
\end{itemize}

The weak-coupling (high-temperature) region of Coulomb systems is described
by the Poisson-Boltzmann (PB) mean-field theory or its linearized version
-- the Debye-H\"uckel theory \cite{Attard02,Andelman06}.
Like-charged colloids always effectively repel one another in the weak-coupling
limit \cite{Gouy10,Chapman13,Chan76,Derjaguin87}.
Two-dimensional Coulomb models are exactly solvable, besides the
high-temperature limit, also at a specific value $\Gamma=2$
of the coupling constant (\ref{coupling}).
The OCP is mappable onto a system of free fermions
\cite{Jancovici81,Alastuey81} while the TCP onto the so-called
Thirring field model \cite{Cornu87,Cornu89}.
The thermodynamics and many-body densities of these 2D Coulomb fluids were
obtained in the bulk as well as semi-infinite and fully finite geometries,
see reviews \cite{Jancovici92,Forrester98}.
The complete thermodynamics and the asymptotic behavior of the charge and
density particle correlation functions are available for the 2D Coulomb gas
even in the whole stability region of the coupling constant $0<\Gamma<2$
via an equivalence with the integrable 2D sine-Gordon field theory 
\cite{Samaj00,Samaj03,Samaj13}.
The strong-coupling (low-temperature) regime, studied mainly for
charged colloid surfaces with counterions only, is controversial.
The leading term of functional approaches based on a virial fugacity expansion
\cite{Moreira00,Moreira01,Netz01}, corresponding to a single-particle theory,
agrees with Monte-Carlo simulations
\cite{Moreira00,Moreira01,Moreira02,Kanduc07}, but higher-order terms fail.
Other approaches based on the creation of classical Wigner crystals on
charged walls by Coulomb particles at zero temperatures
\cite{Shklovskii99,Grosberg02,Samaj11a,Samaj11b} or the idea of
the correlation hole \cite{Nordholm84,Forsman04,Palaia18}
reproduce the leading single-particle theory and imply correction
terms which are in good agreement with Monte-Carlo data.
These strong-coupling theories explain a counter-intuitive effective
attraction of likely-charged plates observed at low enough temperatures,
experimentally \cite{Khan85,Kjellander88,Bloomfield91,Kekicheff93,Dubois98}
as well as by computer simulations \cite{Gulbrand84,Kjellander84,Gronbech97}.

There exists an exact relation known as the contact value theorem
which holds in any spatial dimension.
It was originally derived for planar interfaces carrying a {\em uniform}
surface charge density, with no dielectric jump between the medium
the particles are immersed in and the material of the plates/walls
\cite{Carnie81,Henderson78,Henderson79,Blum81,Wennerstrom82}.
In the geometry of one EDL with planar interface, the contact value theorem
relates the bulk pressure to the particle number density at the interface
and the surface charge density.
In the geometry of two parallel EDLs at a certain distance,
it relates the pressure (the effective force) between EDLs
to the particle number density and the surface charge density
at the interface, separately for each of the two EDLs.
The generalization of the contact theorem to jellium systems
with a volume background charge density \cite{Choquard80,Totsuji81}
involves an additional space integral over the particle charge density.
An attempt to extend the contact value theorem to planar EDLs
with dielectric discontinuity \cite{Carnie81b,Jancovici82} leads to
the appearance of two-point particle correlation functions, while
the usage of field-theoretical techniques \cite{Dean03} induces
an additional Casimir-like term. 
In the framework of the cell model \cite{Levin02,Fuoss51,Marcus55,Deserno01},
the bulk system is modeled by a single charged body of cylindrical or
spherical shape, enclosed together with counterions and salt ions
in a concentric electroneutral Wigner-Seitz cell of similar shape.
The generalisation of the contact value theorem to curved wall's boundaries
occurring in the context of the cell model was made in
\cite{Wennerstrom82,Deserno01,Mallarino15}.

A uniform surface charge density is a crude simplification of real EDLs.
The discreteness of the surface charge density is omnipresent
in bio-interfacial phenomena \cite{Israelachvili92,Leckband93,Walz98}.
Initial theoretical studies of the surface charge modulation were based
on liquid-state theory \cite{Chan80,Kjellander88b} and a combination
of perturbation techniques with Monte-Carlo simulations in the weak-coupling
PB \cite{Lukatsky02a,Henle04} and strong-coupling \cite{Fleck05} regimes.
Another approach used an expansion in the Fourier modes of the surface
charge modulations for salty solutions \cite{Travesset05}.
The obtained results indicate an increase of the counterion density close to
surfaces with modulated charge density (in comparison with those carrying
the uniform charge density of the same mean value) and a reduction
of the pressure between two parallel interfaces \cite{Lukatsky02b,Khan05}.
In the 3D case of one EDL with the surface charge modulated along
one direction only, exact PB solutions were constructed in an inverse
way by exploring the general result for the 2D Liouville equation
(models with counterions only) and the two-soliton solutions of the
2D sinh-Gordon equation (models with salt added) \cite{Samaj19}.

The partition function and many-particle densities of 2D one-component
systems with the coupling constant $\Gamma=2\gamma$ ($\gamma$ is a positive
integer) can be expressed in terms of an anticommuting-field theory
defined on a one-dimensional chain of sites \cite{Samaj95,Samaj04a}.
This mapping was used recently \cite{Samaj22} to derive, under certain
conditions on the matrix of interaction strengths among anticommuting
variables, exact formulas for the density profile and the pressure
for special interfaces with modulated line charge densities at
the free-fermion coupling $\Gamma=2$.
As a by-product of special transformations of anticommuting field variables
leaving the composite form of their action invariant, the contact value
theorem was generalised to interfaces with modulated line charge densities
for any coupling constant $\Gamma=2\gamma$ with $\gamma$ a positive integer
(and therefore by analytic continuation to all real $\Gamma$s in the fluid
region), see equations (4.5) and (5.7), (5.8) of Ref. \cite{Samaj22}
for the geometries of one EDL and two parallel EDLs, respectively.
The generalisation of the contact value theorem to 2D one-component
systems with modulated line charge densities seemed to be related to
special techniques available only in 2D.
However, being motivated by the exact 2D result, we show in this article
that the generalisation can be derived alternatively based on balance of
forces exerted on interface(s) by mobile charges.
This enables us to extend the contact value theorem for modulated surface
charge densities from the special case of the 2D one-component plasma  
to multi-component Coulomb fluids in any spatial dimension.

The paper is organised as follows.
The case of one EDL with modulated surface charge density is the subject
of section \ref{Sec2}.
Basic formalism, the notation and the main 2D result of reference
\cite{Samaj22} are presented in section \ref{Sec21}.
The derivation of the contact value theorem for multi-component Coulomb fluids
in any spatial dimensions, based on balance of forces exerted on interface
by mobile charges, is given in section \ref{Sec22}.
The case of two parallel EDLs with modulated surface charge densities
is discussed in section \ref{Sec3}.
As before, basic formalism and the derivation of the contact value theorem
are presented in sections \ref{Sec31} and \ref{Sec32}, respectively.
Section \ref{Sec4} deals with the analysis of relevance of particular
terms in the contact value relation within the framework of a 2D
exactly solvable model.
A brief recapitulation and concluding remarks are given in section \ref{Sec5}. 

\renewcommand{\theequation}{2.\arabic{equation}}
\setcounter{equation}{0}

\section{Geometry of one EDL} \label{Sec2}

\subsection{Basic formalism and notation} \label{Sec21}
\begin{figure}[tbp]
\centering
\includegraphics[clip,width=0.7\textwidth]{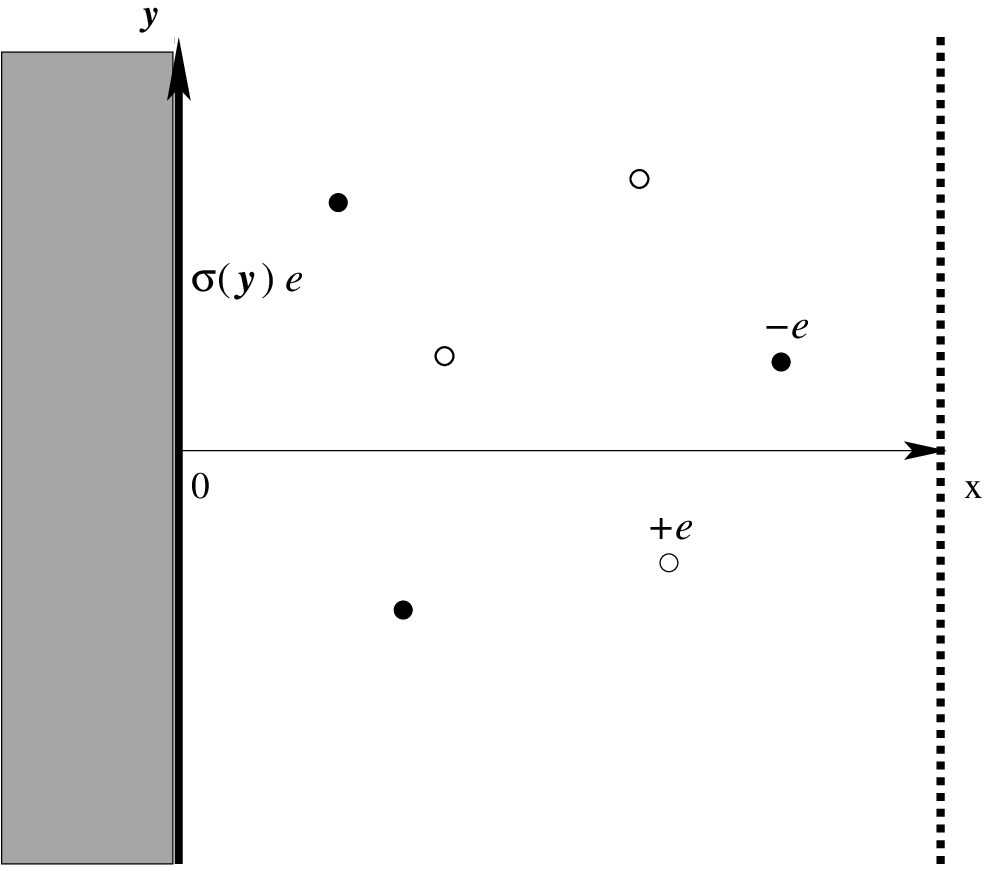}
\caption{The geometry of one EDL.
The wall in the half-space $x<0$ mimics the interior of a colloid,
pointlike particles with $\pm e$ charges of the symmetric TCP move
in the complementary half-space $x>0$.
The $(\nu-1)$-dimensional interface of points ${\bf y}=(y_1,\ldots,y_{\nu-1})$,
located at $x=0$, carries a modulated surface charge density
$\sigma({\bf y}) e$.
There is another interface without any surface charge at $x\to\infty$,
denoted by the dashed line.}
\label{fig1}
\end{figure}

The geometry of one EDL with modulated surface charge density is presented
in figure \ref{fig1}.
We consider an infinite $\nu$-dimensional Euclidean space of points
${\bf r}=(x,{\bf y})$ where Cartesian coordinates $x\in \mathbb{R}$ and
${\bf y}=(y_1,\ldots,y_{\nu-1})\in \mathbb{R}^{\nu-1}$.
The dielectric wall in the half-space $x<0$ mimics the interior of a colloid,
pointlike particles move in the complementary half-space
$\Lambda=\{x>0,{\bf y}\}$.
The $\pm e$ charges of particles means that the pictured system is the
symmetric TCP, the OCP contains particles of the same charge (say $-e$).
The $(\nu-1)$-dimensional interface of points
$\partial\Lambda=\{ {\bf r}=(x=0,{\bf y})\}$ carries a modulated
surface charge density $\sigma({\bf y}) e$.
In real experiments, $\sigma({\bf y})$ is a periodic
function of ${\bf y}$ and it is finite at every point
${\bf y}\in\partial\Lambda$.
The surface of the interface is taken infinite,
$\vert\partial\Lambda\vert \to \infty$.
Dielectric constants of the wall $\varepsilon_w$ and of the medium the
particles are immersed in $\varepsilon$ are taken to be the same,
$\varepsilon_w = \varepsilon = 1$, so that there are no dielectric
image charges.

\subsubsection{OCP}
For the OCP with counterions of charge $-e$, there are $N$ particles
moving in the half-space domain $\Lambda$.
The condition of the overall electroneutrality reads as
\begin{equation} \label{electro1}
N (-e) + \int_{\partial\Lambda} {\rm d}{\bf y}\, \sigma({\bf y}) e = 0 .
\end{equation}

For a given configuration of $N$ charges
$\{ {\bf r}_1,{\bf r}_2,\ldots,{\bf r}_N \}$, the microscopic density
of particles at point ${\bf r}\in\Lambda$ is defined by
$\hat{n}({\bf r}) = \sum_{j=1}^N \delta({\bf r}-{\bf r}_j)$.
The averaged particle number density at point ${\bf r}\in\Lambda$ is given by
\begin{equation}
n({\bf r}) = \langle \hat{n}({\bf r}) \rangle ,
\end{equation}  
where $\langle\cdots\rangle$ denotes the statistical average over the
canonical ensemble at the inverse temperature $\beta$.
The total number of particles is equal to
\begin{equation} \label{N}
N = \int_{\Lambda} {\rm d}{\bf r}\, n({\bf r}) .
\end{equation}
As there is a finite number of particles per unit surface of the interface,
the density must vanish at large distances from the interface,
\begin{equation} \label{asymp}
\lim_{x\to\infty} n({\bf r}) = 0 .  
\end{equation}
This automatically means that the bulk pressure (the derivative of
the free energy with respect to the volume) $P$ vanishes as well.

The microscopic charge density is defined by
$\hat{\rho}({\bf r}) = \sum_{j=1}^N (-e) \delta({\bf r}-{\bf r}_j)$
and the corresponding averaged charge density
$\rho({\bf r}) = \langle \hat{\rho}({\bf r}) \rangle$.
Its relation to the averaged particle density is trivial for the OCP:
\begin{equation}
\rho({\bf r}) = - e n({\bf r}) .  
\end{equation}
The electroneutrality condition (\ref{electro1}) together with
equation (\ref{N}) imply that
\begin{equation} \label{electro2}
\int_{\Lambda} {\rm d}{\bf r}\, \rho({\bf r}) +
\int_{\partial\Lambda} {\rm d}{\bf y}\, \sigma({\bf y}) e = 0 .
\end{equation}

The theory simplifies itself substantially when the surface charge
density is uniform, $\sigma({\bf y}) = \sigma$.
The averaged particle and charge densities then depend only on the coordinate
$x$ perpendicular to the interface, i.e., $n({\bf r}) = n(x)$
and $\rho({\bf r}) = \rho(x)$.
The electroneutrality condition (\ref{electro2}) and the asymptotic
relation (\ref{asymp}) then become
\begin{equation}
\int_0^{\infty} {\rm d}x\, n(x) = \sigma , \qquad
\lim_{x\to\infty} n(x) = 0 .    
\end{equation}  
The contact value theorem
\cite{Carnie81,Henderson78,Henderson79,Blum81,Wennerstrom82}
fixes the density of counterions at the interface as follows
\begin{equation} \label{cvtOCP}
n(0) = \frac{1}{2} s_{\nu} \beta e^2 \sigma^2 ,
\end{equation}
keeping in mind that the bulk pressure $P=0$ for the considered OCP.
This contact value relation was checked on the exact solutions of
the OCP in the weak-coupling PB limit in any dimension \cite{Andelman06} and
at the free fermion coupling $\Gamma=2$ in 2D \cite{Jancovici84,Samaj15}. 

\subsubsection{TCP}
Let the TCP be composed of $N_+$ particles with charge $+e$ and $N_-$ particles
with charge $-e$, the total number of particles $N=N_++N_-$.
The electroneutrality condition reads as
\begin{equation} \label{electro3}
N_+ - N_- = \int_{\partial\Lambda} {\rm d}{\bf y}\, \sigma({\bf y}) .
\end{equation}
For a configuration of $N_+$ particles with charge $+e$,
$\{ {\bf r}^+_1,{\bf r}^+_2,\ldots,{\bf r}^+_{N_+} \}$, the microscopic density
of $+e$ particles at point ${\bf r}\in\Lambda$ is 
$\hat{n}_+({\bf r}) = \sum_{j=1}^{N_+} \delta({\bf r}-{\bf r}^+_j)$ and
the averaged number density of $+e$ particles at point ${\bf r}\in\Lambda$
is $n_+({\bf r}) = \langle \hat{n}_+({\bf r}) \rangle$.
Analogously, for a configuration of $N_-$ particles with charge $-e$,
$\{ {\bf r}^-_1,{\bf r}^-_2,\ldots,{\bf r}^-_{N_-} \}$, the microscopic density
of $-e$ particles at point ${\bf r}\in\Lambda$ is 
$\hat{n}_-({\bf r}) = \sum_{j=1}^{N_-} \delta({\bf r}-{\bf r}^-_j)$ and
the averaged number density of $-e$ particles at point ${\bf r}\in\Lambda$
is $n_-({\bf r}) = \langle \hat{n}_-({\bf r}) \rangle$.
The total averaged number density of particles $n({\bf r})$ and charge density
$\rho({\bf r})$ are given by
\begin{equation}
n({\bf r}) = n_+({\bf r}) + n_-({\bf r}) , \qquad  
\rho({\bf r}) = e \left[ n_+({\bf r}) - n_-({\bf r}) \right] .
\end{equation}  
As there is a finite charge per unit surface of the interface,
the charge density induced by particles must vanish at asymptotically
large distances from the interface,
\begin{equation} \label{asymptot1}
\lim_{x\to\infty} \rho({\bf r}) = 0 .
\end{equation}
On the other hand, the particle number density is, in general, nonzero
and constant (due to the charge screening, not touched by the surface charge
density) at asymptotically large distances from the interface,
\begin{equation} \label{asymptot2}
\lim_{x\to\infty} n({\bf r}) = n ,
\end{equation}
where the bulk density $n$ is controlled by the chemical potential.
The electroneutrality condition (\ref{electro3}) can be written as
the previous one (\ref{electro2}).

If the surface charge density is uniform, $\sigma({\bf y}) = \sigma$,
the averaged particle and charge densities depend only on the
$x$-coordinate.
The asymptotic relations (\ref{asymptot1}) and (\ref{asymptot2})
then become
\begin{equation}
\lim_{x\to\infty} \rho(x) = 0 , \qquad
\lim_{x\to\infty} n(x) = n .    
\end{equation}  
The contact value theorem
\cite{Carnie81,Henderson78,Henderson79,Blum81,Wennerstrom82}
takes the form
\begin{equation} \label{cvtTCP}
\beta P = n(0) - \frac{1}{2} s_{\nu} \beta e^2 \sigma^2 ,
\end{equation}
where the bulk pressure $P$, corresponding to the particle
density $n$, includes electrostatic as well as non-electrostatic
interactions like the Lennard-Jones interaction, the excluded volume effects
\cite{Frydel12}, etc.
Notice that the contact theorem for the OCP (\ref{cvtOCP}), characterized
by $P=0$, is in fact identical to the one for the TCP (\ref{cvtTCP}).
The relation (\ref{cvtTCP}) was checked by using the exact solutions of
the TCP in the weak-coupling PB limit in any dimension \cite{Andelman06}
and at the coupling $\Gamma=2$ in 2D \cite{Cornu87,Cornu89}. 

\subsection{Balance of forces} \label{Sec22}
The contact value theorem for one interface with modulated surface charge
density can be obtained from the balance of the forces exerted to the wall
surface.
The total force along the $x$-direction (perpendicular to the interface) 
consists of three components.

The particles with the density $n(0,{\bf y})$ at the interface
$\partial\Lambda$ push on the wall by the force
\begin{equation}
F^x_1 = - \frac{1}{\beta} \int_{\partial\Lambda} {\rm d}{\bf y}\, n(0,{\bf y}) ,
\end{equation}
oriented to the left along the $x$-axis in figure \ref{fig1}.

The charged particles inside the domain $\Lambda$ induce at the point
$(0,{\bf y}')\in \partial\Lambda$ the electric field
\begin{equation}
E^x_2(0,{\bf y}') = - \int_{\Lambda} {\rm d}{\bf r}\, \rho({\bf r})
\frac{\partial}{\partial (-x)} v(x,\vert {\bf y}-{\bf y}'\vert) .
\end{equation}  
The corresponding force exerted on the surface charge density reads as
\begin{eqnarray}
F^x_2 & = & \int_{\partial\Lambda} {\rm d}{\bf y}'\, \sigma({\bf y}') e
E^x_2(0,{\bf y}') \nonumber \\
& = & e \int_{\partial\Lambda} {\rm d}{\bf y}'\, \sigma({\bf y}')
\int_{\Lambda} {\rm d}{\bf r}\, \rho({\bf r})
\frac{\partial}{\partial x} v(x,\vert {\bf y}-{\bf y}'\vert) . \label{F2x}
\end{eqnarray}  

There is another wall at $x\to\infty$ with no surface charge density
and of the same surface $\vert\partial\Lambda\vert$ as the one at $x=0$,
see the dashed line in figure \ref{fig1}.
The particles of bulk density $n$ (controlled by the chemical potential)
push on this wall by the force
\begin{equation}
F^x_3 = P \vert\partial\Lambda\vert ,
\end{equation}  
where $P$ is the bulk pressure corresponding to the particle density $n$.

The total force acting on the interface $\partial\Lambda$ and the one at
infinity must be zero in thermal equilibrium, i.e.,
\begin{equation} \label{balanceforces}
F^x_1 + F^x_2 + F^x_3 = 0 .
\end{equation}  
Introducing the mean particle density at the interface
\begin{equation}
\bar{n}(0) \equiv \frac{1}{\vert\partial\Lambda\vert} \int_{\partial\Lambda}
{\rm d}{\bf y}\, n(0,{\bf y}) ,     
\end{equation}
the balance of forces (\ref{balanceforces}) implies the relation
\begin{equation} \label{final1}
\beta P = \bar{n}(0) - \frac{\beta e}{\vert\partial\Lambda\vert}
\int_{\partial\Lambda} {\rm d}{\bf y}'\, \sigma({\bf y}')
\int_{\Lambda} {\rm d}{\bf r}\, \rho({\bf r})
\frac{\partial}{\partial x} v(x,\vert {\bf y}-{\bf y}'\vert) .   
\end{equation}

This relation can be adapted further by considering the point-dependent
deviations of the surface charge density from its mean value $\bar{\sigma}$,
\begin{equation} \label{deviation}
\sigma({\bf y}) = \bar{\sigma} + \delta\sigma({\bf y}) , \qquad  
\bar{\sigma} \equiv \frac{1}{\vert\partial\Lambda\vert} \int_{\partial\Lambda}
{\rm d}{\bf y}\, \sigma({\bf y}) .
\end{equation}
The mean value $\bar{\sigma}$ is mathematically well
defined since the surface charge density $\sigma({\bf y})$ is assumed
to be finite at every point ${\bf y}\in\partial\Lambda$.
The surface integral over deviations must vanish by definition,
\begin{equation}
\int_{\partial\Lambda} {\rm d}{\bf y}\, \delta\sigma({\bf y}) = 0 .
\end{equation}  
Inserting the decomposition (\ref{deviation}) into (\ref{final1}), one gets
\begin{eqnarray}
\beta P & = & \bar{n}(0) -
\frac{\beta e \bar{\sigma}}{\vert\partial\Lambda\vert}
\int_{\partial\Lambda} {\rm d}{\bf y}' \int_{\Lambda} {\rm d}{\bf r}\, \rho({\bf r})
\frac{\partial}{\partial x} v(x,\vert {\bf y}-{\bf y}'\vert) \nonumber \\ & &
- \frac{\beta e}{\vert\partial\Lambda\vert}
\int_{\partial\Lambda} {\rm d}{\bf y}'\, \delta\sigma({\bf y}')
\int_{\Lambda} {\rm d}{\bf r}\, \rho({\bf r})
\frac{\partial}{\partial x} v(x,\vert {\bf y}-{\bf y}'\vert) . \label{final2}  
\end{eqnarray}

According to Fubini's theorem, the order of integration of
an absolutely integrable function can be exchanged \cite{Fubini1907}.
Since
\begin{equation}
\frac{\partial}{\partial x} v(x,\vert {\bf y}-{\bf y}'\vert)
= - \frac{x}{(x^2+\vert {\bf y}-{\bf y'}\vert^2)^{\nu/2}}   
\end{equation}
and the charge density $\rho({\bf r})$ is finite at every point
${\bf r}\in \Lambda$, one can interchange the order of integrations over
${\bf y}'$ and ${\bf r}$ in the first integral on the rhs of equation
(\ref{final2}) and concentrate on the integral
\begin{equation} \label{integral}
\int_{\partial\Lambda} {\rm d}{\bf y}'
\frac{\partial}{\partial x} v(x,\vert {\bf y}-{\bf y}'\vert)
= \int_{\partial\Lambda} {\rm d}{\bf y}'\,
\frac{\partial}{\partial x} v(x,\vert {\bf y}'\vert) ,
\end{equation}
where the substitution ${\bf y}'-{\bf y}\to {\bf y}'$ was made
within an infinite domain's boundary $\partial\Lambda$.
Using the radial coordinate system on the $(\nu-1)$-dimensional
interface $\partial\Lambda$, the integral (\ref{integral}) can be expressed as
\begin{equation}
- \int_0^{\infty} {\rm d}y'\, s_{\nu-1} {y'}^{\nu-2} \frac{x}{(x^2+y'^2)^{\nu/2}}
= - s_{\nu-1} \frac{\sqrt{\pi}
\Gamma\left(\frac{\nu-1}{2}\right)}{2\Gamma\left(\frac{\nu}{2}\right)}
= - \frac{1}{2} s_{\nu} .
\end{equation}
The first integral on the rhs of equation (\ref{final2}) can be thus written as
\begin{equation}
\int_{\partial\Lambda} {\rm d}{\bf y}' \int_{\Lambda} {\rm d}{\bf r}\, \rho({\bf r})
\frac{\partial}{\partial x} v(x,\vert {\bf y}-{\bf y}'\vert) 
= - \frac{1}{2} s_{\nu} \int_{\Lambda} {\rm d}{\bf r}\, \rho({\bf r}) .
\end{equation}
Finally, since the electroneutrality condition (\ref{electro2}) implies that
\begin{equation} \label{elect} 
\int_{\Lambda} {\rm d}{\bf r}\, \rho({\bf r}) = 
- \int_{\partial\Lambda} {\rm d}{\bf y}\, \sigma({\bf y}) e
= - \bar{\sigma} e \vert\partial\Lambda\vert ,
\end{equation}
equation (\ref{final2}) simplifies itself to
\begin{eqnarray}
\beta P & = & \bar{n}(0) - \frac{1}{2} s_{\nu} \beta e^2 \bar{\sigma}^2
\nonumber \\ & & - \frac{\beta e}{\vert\partial\Lambda\vert}
\int_{\partial\Lambda} {\rm d}{\bf y}'\, \delta\sigma({\bf y}')
\int_{\Lambda} {\rm d}{\bf r}\, \rho({\bf r})
\frac{\partial}{\partial x} v(x,\vert {\bf y}-{\bf y}'\vert) . \label{final3}  
\end{eqnarray}
We see that the modulation of the surface charge density induces
into the contact value theorem the charge density of particles inside
the whole domain $\Lambda$.
In the uniform case $\delta\sigma({\bf y})=0$ with $\bar{n}(0)=n(x=0)$,
the previous contact value relation (\ref{cvtTCP}) is reproduced.
Note that the derivation of the contact value theorem
(\ref{final3}) was based on the consideration of balance of electrostatic
forces, the non-electrostatic forces contribute only to the bulk
pressure $P$.

As for the force components along directions ${\bf y}$
parallel to the interface, it is first necessary to specify boundary
conditions at $\pm$infinity.
In analogy with the previous work \cite{Samaj22}, let us consider
the periodic boundary conditions for which only the component of type
(\ref{F2x}) survives:
\begin{eqnarray}
F^{y_j}_2 & = & \int_{\partial\Lambda} {\rm d}{\bf y}'\, \sigma({\bf y}') e
E^{y_j}_2(0,{\bf y}') \nonumber \\
& = & e \int_{\partial\Lambda} {\rm d}{\bf y}'\, \sigma({\bf y}')
\int_{\Lambda} {\rm d}{\bf r}\, \rho({\bf r})
\frac{\partial}{\partial y_j} v(x,\vert {\bf y}-{\bf y}'\vert) \label{F2y}
\end{eqnarray}  
$(j=1,2,\ldots,\nu-1)$.
Using here the decomposition (\ref{deviation}), changing the order
of integrations attached to the term $\bar{\sigma}$ and setting the force
in equilibrium equal to zero, one obtains the equality
\begin{equation}
\int_{\partial\Lambda} {\rm d}{\bf y}'\, \delta\sigma({\bf y}')
\int_{\Lambda} {\rm d}{\bf r}\, \rho({\bf r})
\frac{\partial}{\partial y_j} v(x,\vert {\bf y}-{\bf y}'\vert) = 0 ,
\quad j=1,2,\ldots,\nu-1.
\end{equation}
In 2D, this relation was derived by using symmetry transformations of
anticommuting variables, see equation (4.8) in Ref. \cite{Samaj22}.

\renewcommand{\theequation}{3.\arabic{equation}}
\setcounter{equation}{0}

\section{Geometry of two parallel EDLs} \label{Sec3}

\subsection{Basic formalism and notation} \label{Sec31}
\begin{figure}[tbp]
\centering
\includegraphics[clip,width=0.7\textwidth]{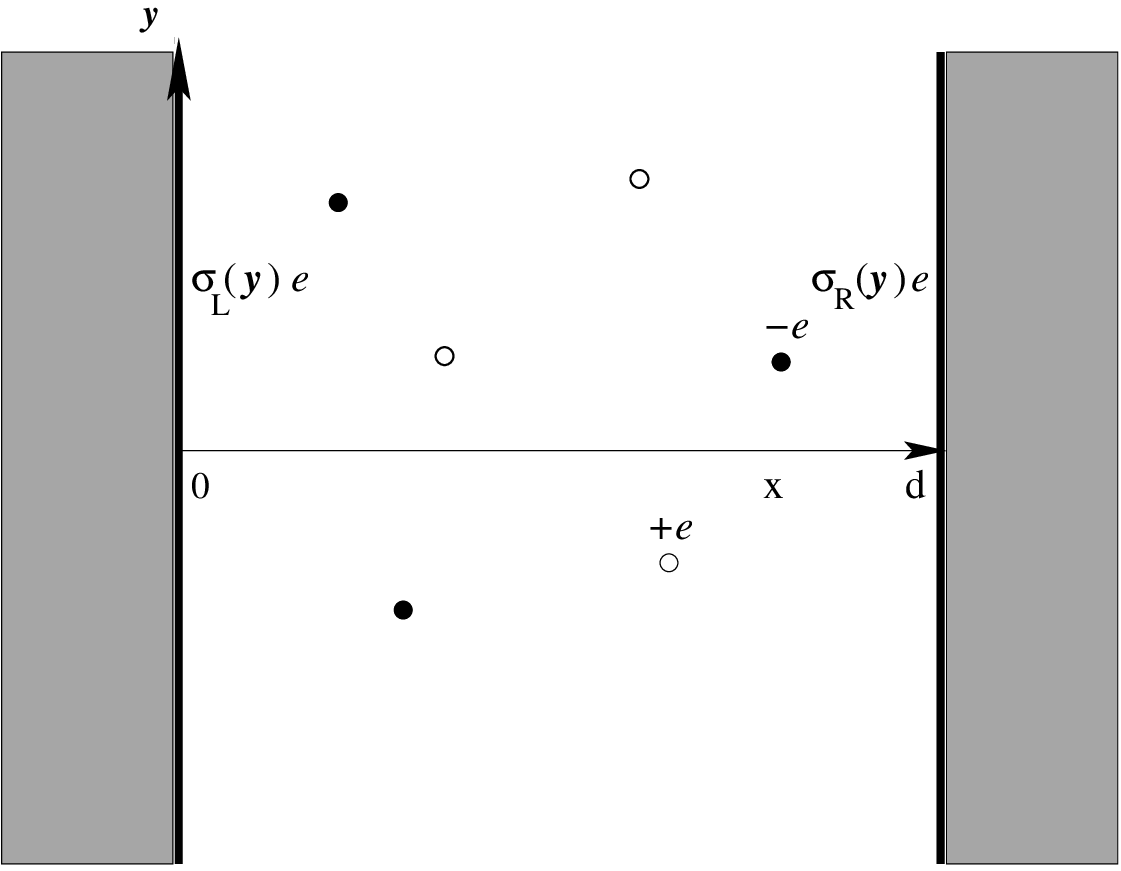}
\caption{The geometry of two parallel EDLs.
The wall in the half-space $x<0$, with interface $\partial\Lambda_L$ at $x=0$,
carries a modulated surface charge density $\sigma_L({\bf y}) e$.  
The wall in the half-space $x>d$, with interface $\partial\Lambda_R$ at $x=d$,
carries $\sigma_R({\bf y}) e$.  
Pointlike particles of the symmetric TCP with $\pm e$ charges move 
inside the domain
$\Lambda=\{ {\bf r}=(x,{\bf y}), 0<x<d, {\bf y}\in \mathbb{R}^{\nu-1} \}$.}
\label{fig2}
\end{figure}

The geometry of two parallel interfaces at distance $d$ is pictured
in figure \ref{fig2}.
The ``left'' wall in the half-space $x<0$ with interface
$\partial\Lambda_L$ at $x=0$ carries a modulated surface charge
density $\sigma_L({\bf y}) e\ge 0$.  
There is another ``right'' wall in the half-space $x>d$, with interface
$\partial\Lambda_R$ at $x=d$, which carries $\sigma_R({\bf y}) e\ge 0$.
The surfaces of the interfaces are taken to be the same and going to infinity,
\begin{equation}
\vert \partial\Lambda_L\vert = \vert \partial\Lambda_R\vert
= \vert \partial\Lambda\vert \to \infty .
\end{equation}  
$N$ pointlike particles of the symmetric TCP with $\pm e$ charges move 
inside the domain
$\Lambda=\{ {\bf r}=(x,{\bf y}), 0<x<d, {\bf y}\in \mathbb{R}^{\nu-1} \}$.
A potential OCP contains only $-e$ charges.

Introducing the averaged particle number $n({\bf r};d)$ and
charge $\rho({\bf r};d)$ densities, the condition of the overall
electroneutrality reads as
\begin{equation} \label{electro4}
\int_{\Lambda} {\rm d}{\bf r}\, \rho({\bf r};d)
+ \int_{\partial\Lambda_L} {\rm d}{\bf y}\, \sigma_L({\bf y}) e
+ \int_{\partial\Lambda_R} {\rm d}{\bf y}\, \sigma_R({\bf y}) e = 0 .
\end{equation}

When the surface charge densities are uniform, i.e.,
$\sigma_L({\bf y}) = \sigma_L$ and $\sigma_R({\bf y}) = \sigma_R$,
the averaged particle number and charge densities depend only on
the $x$-coordinate, $n({\bf r};d) = n(x;d)$ and $\rho({\bf r};d) = \rho(x;d)$. 
The contact value theorem 
\cite{Carnie81,Henderson78,Henderson79,Blum81,Wennerstrom82}
relates the pressure of the Coulomb fluid between EDLs $P$ and the
contact particle number densities on both interfaces as follows
\begin{eqnarray} \label{cvt}
\beta P(d) & = & n(0;d) - \frac{1}{2} s_{\nu} \beta e^2 \sigma_L^2 \nonumber \\
& = & n(d;d) - \frac{1}{2} s_{\nu} \beta e^2 \sigma_R^2 .  
\end{eqnarray}
In the limit of infinite distance between the two EDLs $d\to\infty$,
the system decomposes itself onto two separate (noninteracting) EDLs with
the bulk pressure
\begin{equation} \label{Pd}
\lim_{d\to\infty}P(d) = P ;
\end{equation}
the relations (\ref{cvt}) then correspond to two independent contact value
theorems for one EDL of type (\ref{cvtTCP}).
Taking the left and right walls as plates of a finite thickness
(as it is in the case of two big colloids), the effective force
per unit surface between the plates mediated by the Coulomb fluid in between
is equal to $P(d)-P$.
In view of relation (\ref{Pd}), this force goes to 0 at $d\to\infty$ as
it should be.
The positive (negative) value of the pressure $P(d)-P>0$ $(P(d)-P<0)$ means
the repulsion (attraction) of the interfaces.

\subsection{Balance of forces} \label{Sec32}
The force acting on the left interface is balanced completely by
the opposite force acting on the right interface, so that the total force
on the Coulomb system in thermal equilibrium vanishes as it should be.
Note that all considered forces act along the $x$-axis. 

Let us first consider the left wall with the interface $\partial\Lambda_L$.
In analogy with section \ref{Sec22}, the particles with the density
$n(0,{\bf y})$ push on $\partial\Lambda_L$ by the force
\begin{equation}
F^x_1 = - \frac{1}{\beta} \int_{\partial\Lambda} {\rm d}{\bf y}\, n(0,{\bf y}) .
\end{equation}
The force induced by the charged particles inside the domain $\Lambda$
reads as
\begin{equation}
F^x_2 = e \int_{\partial\Lambda} {\rm d}{\bf y}'\, \sigma_L({\bf y}')
\int_{\Lambda} {\rm d}{\bf r}\, \rho({\bf r})
\frac{\partial}{\partial x} v(x,\vert {\bf y}-{\bf y}'\vert) . 
\end{equation}  
The particles push on the opposite wall with interface $\partial\Lambda_R$
by the force
\begin{equation}
F^x_3 = P(d) \vert\partial\Lambda\vert .
\end{equation}  
Finally, the force exerted by the charged interface $\partial\Lambda_R$
on the one $\partial\Lambda_L$ is given by
\begin{equation}
F^x_4 = e^2 \int_{\partial\Lambda_L} {\rm d}{\bf y}
\int_{\partial\Lambda_R} {\rm d}{\bf y}'\, \sigma_L({\bf y})
\frac{\partial}{\partial d} v(d,\vert {\bf y}-{\bf y}'\vert)
\sigma_R({\bf y}') .
\end{equation}  

The total force acting on the two interfaces must be zero,
\begin{equation} \label{balanceforces2}
F^x_1 + F^x_2 + F^x_3 + F^x_4 = 0 .
\end{equation}  
In terms of the mean particle density at the interface $\partial\Lambda_L$
\begin{equation} \label{n0}
\bar{n}(0) = \frac{1}{\vert\partial\Lambda\vert} \int_{\partial\Lambda_L}
{\rm d}{\bf y}\, n(0,{\bf y}) ,     
\end{equation}
the balance of forces (\ref{balanceforces2}) leads to the relation
\begin{eqnarray} 
\beta P(d) & = & \bar{n}(0) - \frac{\beta e}{\vert\partial\Lambda\vert}
\int_{\partial\Lambda_L} {\rm d}{\bf y}'\, \sigma_L({\bf y}')
\int_{\Lambda} {\rm d}{\bf r}\, \rho({\bf r})
\frac{\partial}{\partial x} v(x,\vert {\bf y}-{\bf y}'\vert) \nonumber \\
& & - \frac{\beta e^2}{\vert\partial\Lambda\vert}
\int_{\partial\Lambda_L} {\rm d}{\bf y}
\int_{\partial\Lambda_R} {\rm d}{\bf y}'\, \sigma_L({\bf y})
\frac{\partial}{\partial d} v(d,\vert {\bf y}-{\bf y}'\vert)
\sigma_R({\bf y}') . \label{final4}
\end{eqnarray}

For each of the interfaces, we introduce point-dependent
deviations of the surface charge density from its mean value,
\begin{equation} \label{deviation1}
\sigma_L({\bf y}) = \bar{\sigma}_L + \delta\sigma_L({\bf y}) , \qquad  
\bar{\sigma}_L = \frac{1}{\vert\partial\Lambda\vert} \int_{\partial\Lambda_L}
{\rm d}{\bf y}\, \sigma_L({\bf y}) ,
\end{equation}
\begin{equation} \label{deviation2}
\sigma_R({\bf y}) = \bar{\sigma}_R + \delta\sigma_R({\bf y}) , \qquad  
\bar{\sigma}_R = \frac{1}{\vert\partial\Lambda\vert} \int_{\partial\Lambda_R}
{\rm d}{\bf y}\, \sigma_R({\bf y}) .
\end{equation}
The surface integral over deviations must vanish for each of the interfaces,
\begin{equation}
\int_{\partial\Lambda_L} {\rm d}{\bf y}\, \delta\sigma_L({\bf y}) = 0 , \qquad
\int_{\partial\Lambda_R} {\rm d}{\bf y}\, \delta\sigma_R({\bf y}) = 0 .  
\end{equation}
Inserting the decompositions (\ref{deviation1}) and (\ref{deviation2})
into (\ref{final4}), we do again the procedure between equations
(\ref{integral}) and (\ref{elect}) of section \ref{Sec22}.
With the aid of the electroneutrality condition (\ref{electro4}), 
the total pressure $P$ is obtained as the sum of two contributions,
\begin{equation} \label{contributions}
\beta P(d) = \beta P_{\rm part}(d) + \beta P_{\rm dev}(d) ,
\end{equation}
where
\begin{eqnarray} 
\beta P_{\rm part}(d) & = & \bar{n}(0) - \frac{1}{2} s_{\nu} \beta e^2
\bar{\sigma}_L^2 \nonumber \\ & &
- \frac{\beta e}{\vert\partial\Lambda\vert}
\int_{\partial\Lambda_L} {\rm d}{\bf y}'\, \delta\sigma_L({\bf y}')
\int_{\Lambda} {\rm d}{\bf r}\, \rho({\bf r})
\frac{\partial}{\partial x} v(x,\vert {\bf y}-{\bf y}'\vert) \label{final51}
\end{eqnarray}
is the ``particle'' part which depends on the mean particle number density
in contact $\bar{n}(0)$ defined in (\ref{n0}) and the profile of the particle
charge density $\rho({\bf r})$ inside the whole domain $\Lambda$, and
\begin{equation} \label{final52}
\beta P_{\rm dev}(d) = - \frac{\beta e^2}{\vert\partial\Lambda\vert}
\int_{\partial\Lambda_L} {\rm d}{\bf y}
\int_{\partial\Lambda_R} {\rm d}{\bf y}'\, \delta\sigma_L({\bf y})
\frac{\partial}{\partial d} v(d,\vert {\bf y}-{\bf y}'\vert)
\delta\sigma_R({\bf y}') 
\end{equation}
is the ``deviation'' part corresponding to the pure Coulomb interaction
of of surface charge deviations on the two interfaces.

Applying the same procedure to the the right interface $\partial\Lambda_R$
results in the split formula (\ref{contributions}) with
the particle part
\begin{eqnarray} 
\beta P_{\rm part}(d) & = & \bar{n}(d)
- \frac{1}{2} s_{\nu} \beta e^2 \bar{\sigma}_R^2
\nonumber \\ & & + \frac{\beta e}{\vert\partial\Lambda\vert}
\int_{\partial\Lambda_R} {\rm d}{\bf y}'\, \delta\sigma_R({\bf y}')
\int_{\Lambda} {\rm d}{\bf r}\ \rho({\bf r})
\frac{\partial}{\partial x} v(d-x,\vert {\bf y}-{\bf y}'\vert) \nonumber \\
\label{final6}
\end{eqnarray}
and the deviation part $\beta P_{\rm dev}(d)$ given by the previous formula
(\ref{final52}).
Here,
\begin{equation}
\bar{n}(d) = \frac{1}{\vert\partial\Lambda\vert} \int_{\partial\Lambda_R}
{\rm d}{\bf y}\, n(d,{\bf y})     
\end{equation}
is the mean particle density in contact with the right interface
$\partial\Lambda_R$. 
Note that in the large-$d$ limit the deviation part
$\beta P_{\rm dev}$ (\ref{final52}) vanishes and one is left with
the pair of independent one-EDL counterparts of type (\ref{final3}).  
 
\renewcommand{\theequation}{4.\arabic{equation}}
\setcounter{equation}{0}

\section{Analysis of an exactly solvable 2D model} \label{Sec4}
As was mentioned in the Introduction, many types of 2D Coulomb fluid
are exactly solvable at the coupling constant $\Gamma=2$.

For the geometry of two parallel lines at distance $d$, carrying
the uniform line charge densities $\sigma_L e$ and $\sigma_R e$
with counterions of charge $-e$ only, the exact formula for the pressure $P$
at $\Gamma=2$ was derived in \cite{Samaj14,Samaj20}:
\begin{equation} \label{pressure1}
\beta P_0(d;\sigma_L,\sigma_R) = \beta P_0(d;\sigma_L) + \beta P_0(d;\sigma_R) ,
\end{equation}
where
\begin{equation} \label{pressure2}
\beta P_0(d;\sigma) = \frac{1}{2\pi d^2} \int_0^{2\pi\sigma d} {\rm d}t\,
\frac{t}{\sinh t} {\rm e}^{-t}
\end{equation}
and the subscript ``0'' in $P_0$ means that the line charge densities are
uniform.
In the case of like-charged lines $0<\sigma_L\le\sigma_R$, the pressure
is always positive, $P_0(d)>0$, i.e., the charged lines repeal each other for
any distance $d$.
The pressure diverges at small distances $d$,
\begin{equation}
\beta P_0(d;\sigma_L,\sigma_R) \mathop{\sim}_{d\to 0}
\frac{\sigma_L+\sigma_R}{d} ,
\end{equation}  
and decays monotonously to 0 from above at asymptotically large $d$, 
\begin{equation} \label{P0d}
\beta P_0(d;\sigma_L,\sigma_R) \mathop{\sim}_{d\to\infty}
\frac{1}{\pi d^2} \int_0^{\infty} {\rm d}t\,
\frac{t}{\sinh t} {\rm e}^{-t} = \frac{\pi}{12} \frac{1}{d^2} .  
\end{equation}
Note that this asymptotic decay is universal, independent of
the line charge densities $\sigma_L$ and $\sigma_R$.

The necessary conditions under which a 2D system with modulated line
charge densities is exactly solvable at $\Gamma=2$ were established
in \cite{Samaj22}.
A version of integrable model is given by the mean line charge densities
$\sigma_L e$ and $\sigma_R e$ and the periodic deviations 
\begin{equation}
\delta\sigma_L(y) = A_L \cos\left( \frac{2\pi}{\lambda}y \right) , \qquad
\delta\sigma_R(y) = A_R \cos\left( \frac{2\pi}{\lambda}y \right) ,  
\end{equation}  
with $\lambda$ being the period and the amplitudes are constrained by
$\vert A_L\vert \le \sigma_L$ and $\vert A_R\vert \le \sigma_R$.
This model is exactly solvable provided that
\begin{equation} \label{condition}
\lambda (\sigma_L + \sigma_R ) \le 1 .
\end{equation}
In terms of the dimensionless pressure
\begin{equation}
\widetilde{P}(d) \equiv \lambda^2 \beta P(d) ,
\end{equation}
the exact solution can be written as the decomposition of type
(\ref{contributions})
\begin{equation} \label{decomp}
\widetilde{P}(d) = \widetilde{P}_{\rm part}(d) + \widetilde{P}_{\rm dev}(d) ,
\end{equation}
with
\begin{eqnarray}
\widetilde{P}_{\rm part}(d) & = & \frac{\partial}{\partial (d/\lambda)}
g\left( d/\lambda,\lambda\sigma_L;\{ \lambda A_L,\lambda A_R\} \right)
\nonumber \\ & & + \frac{\partial}{\partial (d/\lambda)}
g\left( d/\lambda,\lambda\sigma_R;\{ \lambda A_R,\lambda A_L\} \right)
\end{eqnarray}
and
\begin{equation}
\widetilde{P}_{\rm dev}(d) = \pi (\lambda A_L) (\lambda A_R)
{\rm e}^{-2\pi d/\lambda} .
\end{equation}
Here,
\begin{eqnarray}
g\left( d/\lambda,\lambda\sigma;\{ \lambda A,\lambda A'\} \right)
& = & \frac{1}{4\pi} \int_0^{4\pi\lambda\sigma} {\rm d}r\,
\ln\Bigg\{ \int_0^{d/\lambda} {\rm d}x\, {\rm e}^{-r x} \nonumber \\ & & \times
I_0\left(\lambda A {\rm e}^{-2\pi x} + \lambda A' {\rm e}^{-2\pi d/\lambda+2\pi x}
\right) \Bigg\} , \label{g}
\end{eqnarray}  
where $I_0$ denotes the modified Bessel function of the first kind
\cite{Gradshteyn}.

For simplicity, we shall restrict ourselves to the symmetrically charged
lines with the equivalent mean values of the line charge densities
\begin{equation} \label{sigma}
\sigma_L = \sigma_R = \sigma
\end{equation}  
and the equivalent (positive) amplitudes 
\begin{equation}
A_L = A_R = A , \qquad 0<A\le\sigma .
\end{equation}
For this symmetric case
\begin{equation}
\sigma_L(y) = \sigma_R(y) = \sigma
+ A \cos\left( \frac{2\pi}{\lambda} y \right) ,
\end{equation}  
the condition of exact solvability (\ref{condition}) takes the form
\begin{equation}
\lambda \sigma \le \frac{1}{2} .
\end{equation}
Let us consider the extreme value $\lambda\sigma=\frac{1}{2}$ for which
\begin{equation} \label{extreme}
\lambda A \le \lambda \sigma = \frac{1}{2} .
\end{equation}  
The dimensionless pressure is again given by (\ref{decomp}) where
\begin{equation} \label{Ppart}
\widetilde{P}_{\rm part}(d/\lambda,\lambda A)
= 2 \frac{\partial}{\partial (d/\lambda)}
g\left( d/\lambda;\lambda A\right)
\end{equation}
and
\begin{equation} \label{Pint}
\widetilde{P}_{\rm dev}(d/\lambda,\lambda A)
= \pi (\lambda A)^2 {\rm e}^{-2\pi d/\lambda} .
\end{equation}
The $g$-function reads as
\begin{eqnarray}
g\left( d/\lambda;\lambda A\right)
& = & \frac{1}{4\pi} \int_0^{2\pi} {\rm d}r\,
\ln\Bigg\{ \int_0^{d/\lambda} {\rm d}x\, {\rm e}^{-r x} \nonumber \\ & & \times
I_0\left(\lambda A [{\rm e}^{-2\pi x} + {\rm e}^{-2\pi d/\lambda+2\pi x}]
\right) \Bigg\} . \label{gg}
\end{eqnarray}  

The above exact formulas for $\widetilde{P}_{\rm part}(d/\lambda,\lambda A)$
in terms of $g\left( d/\lambda;\lambda A\right)$ are rather complicated and
they can be treated for the given values of the dimensionless amplitudes
$\lambda A$ and distances $d/\lambda$ only numerically.
The analytic treatment is possible in special limits, like for instance
in the limit of small amplitudes $\lambda A$ when the Bessel function in
(\ref{gg}) can be expanded as \cite{Gradshteyn}
\begin{eqnarray}
I_0\left(\lambda A [{\rm e}^{-2\pi x} + {\rm e}^{-2\pi d/\lambda+2\pi x}] \right)
& = & 1 + \frac{(\lambda A)^2}{4}
\Big( {\rm e}^{-4\pi x} + 2 {\rm e}^{-2\pi d/\lambda} \nonumber \\ & &
+ {\rm e}^{-4\pi d/\lambda+4\pi x} \Big) + O\left( (\lambda A)^4\right) . 
\end{eqnarray}
Performing then the integration over the $x$-variable in (\ref{gg}),
the logarithm takes the following argument
\begin{eqnarray}
\ln\Bigg\{ \frac{1-{\rm e}^{-r d/\lambda}}{r} + \frac{(\lambda A)^2}{4}
\Big( \frac{1-{\rm e}^{-(4\pi+r)d/\lambda}}{4\pi+r}
+ 2 {\rm e}^{-2\pi d/\lambda} \frac{1-{\rm e}^{-r d/\lambda}}{r} & & 
\nonumber \\
+ \frac{{\rm e}^{-r d/\lambda}-{\rm e}^{-4\pi d/\lambda}}{4\pi-r} \Big)
+ O\left( (\lambda A)^4\right) \Bigg\} . & &  
\end{eqnarray}
Expanding the logarithm in $\lambda A$ results in
\begin{eqnarray}
g\left( d/\lambda;\lambda A\right)
= \frac{1}{4\pi} \int_0^{2\pi} {\rm d}r\, \ln\left(
\frac{1-{\rm e}^{-r d/\lambda}}{r} \right)
+ \frac{(\lambda A)^2}{4} \frac{1}{4\pi} \int_0^{2\pi} {\rm d}r\,
\nonumber \\ \times 
\left[ \frac{1-{\rm e}^{-(4\pi+r)d/\lambda}}{1-{\rm e}^{-r d/\lambda}} \frac{r}{4\pi+r}
+ 2 {\rm e}^{-2\pi d/\lambda}
+ \frac{{\rm e}^{-r d/\lambda}-{\rm e}^{-4\pi d/\lambda}}{1-{\rm e}^{-r d/\lambda}}
\frac{r}{4\pi-r} \right] \nonumber \\
+ O\left( (\lambda A)^4\right) . \label{ggg}
\end{eqnarray}

The first term on the rhs of this equation is related to the dimensionless
pressure
$\widetilde{P}_0(d;\sigma,\sigma) \equiv \lambda^2 \beta P_0(d,\sigma,\sigma)$
with uniform line charge densities $\sigma_L=\sigma_R=1/(2\lambda)$, see
formulas (\ref{pressure1}) and (\ref{pressure2}), as follows
\begin{equation}
\widetilde{P}_0\left( d;\sigma_L=\frac{1}{2\lambda},
\sigma_R=\frac{1}{2\lambda}\right)
= 2 \frac{\partial}{\partial(d/\lambda)}
\frac{1}{4\pi} \int_0^{2\pi} {\rm d}r\, \ln\left(
\frac{1-{\rm e}^{-r d/\lambda}}{r} \right) .
\end{equation}

The second term on the rhs of equation (\ref{ggg}), proportional to
$(\lambda A)^2$, can be treated analytically in two limits
$d/\lambda\to 0$ and $d/\lambda\to\infty$.
\begin{itemize}
\item  
In the limit $d/\lambda\to 0$, one simply Taylor expands the function under
integration over $r$ in powers of $(d/\lambda)$ and then integrate over $r$,
with the result
\begin{equation}
(\lambda A)^2 \left[ \frac{1}{2} - \frac{\pi d}{\lambda} +
\frac{7}{6} \left( \frac{\pi d}{\lambda} \right)^2 -
\left( \frac{\pi d}{\lambda} \right)^3 +
O\left( \left( \pi d/\lambda \right)^4\right) \right] .    
\end{equation}  
With regard to the relation (\ref{Ppart}), one can write
\begin{eqnarray}
\widetilde{P}_{\rm part}(d/\lambda,\lambda A) & = &
\widetilde{P}_0\left( d;\sigma_L=\frac{1}{2\lambda},
\sigma_R=\frac{1}{2\lambda}\right) + (\lambda A)^2
\Bigg[ -2\pi \nonumber \\ & & + \frac{14\pi}{3} \frac{\pi d}{\lambda}
- 6\pi \left( \frac{\pi d}{\lambda} \right)^2 +
O\left( \left( \pi d/\lambda\right)^3\right) \Bigg] . 
\end{eqnarray}
Comparing the leading term $-2\pi (\lambda A)^2$ to the one
$\pi (\lambda A)^2$ of the line charge interaction pressure
$\widetilde{P}_{\rm dev}(d/\lambda,\lambda A)$ in (\ref{Pint})
we see that it is twice larger by the amplitude
and has a minus sign, in agreement with the hypothesis that
the modulation of the line charge densities diminish the pressure
between the two lines.

\item
In the limit $d/\lambda\to\infty$, one can neglect certain terms
under integration over $r$ vanishing exponentially in this limit, to get
\begin{equation}
\frac{(\lambda A)^2}{4} \frac{1}{4\pi} \int_0^{2\pi} {\rm d}r\,
\left\{ \frac{-r}{4\pi-r} + \frac{1}{1-{\rm e}^{-r d/\lambda}}
\left[ \frac{r}{4\pi+r} + \frac{r}{4\pi-r} \right] \right\} .
\end{equation}
This expression can be further manipulated as follows
\begin{equation}
\frac{(\lambda A)^2}{4} \left[ \left( \frac{1}{2} - \ln 2 \right)
+ \frac{1}{4\pi} \int_0^{2\pi} {\rm d}r\,
\sum_{n=1}^{\infty} {\rm e}^{- n r d/\lambda}
\frac{8\pi r}{(4\pi)^2-r^2} \right] .
\end{equation}
Since it holds
\begin{equation}
\int_0^{2\pi} {\rm d}r\, {\rm e}^{- n r d/\lambda} \frac{2 r}{(4\pi)^2-r^2} 
\mathop{\sim}_{d/\lambda\to\infty} \frac{1}{8\pi^2 n^2 (d/\lambda)^2}
+ O\left( \frac{1}{(d/\lambda)^3} \right) ,
\end{equation}
one ends up with
\begin{equation}
(\lambda A)^2 \left[ \frac{1}{4} \left( \frac{1}{2} - \ln 2 \right)
+ \frac{\zeta(2)}{32\pi^2 (d/\lambda)^2} \right] ,
\end{equation}
where $\zeta(2)$ is the Riemann zeta function at point 2 \cite{Gradshteyn},
\begin{equation}
\zeta(2) = \sum_{n=1}^{\infty} \frac{1}{n^2} = \frac{\pi^2}{6} .
\end{equation}  
Finally, using (\ref{Ppart}) one arrives at
\begin{eqnarray}
\widetilde{P}_{\rm part}(d/\lambda,\lambda A) & = &
\widetilde{P}_0\left( d;\sigma_L=\frac{1}{2\lambda},
\sigma_R=\frac{1}{2\lambda}\right) - (\lambda A)^2 \frac{1}{48}
\frac{1}{(d/\lambda)^3} \nonumber \\ & &
+ O\left( \frac{1}{(d/\lambda)^4} \right) .
\end{eqnarray}
Since the deviation part of the pressure (\ref{Pint}) decays
exponentially at large distances $d/\lambda$, the above obtained long-ranged
term dominates and its negative sign indicates the diminution of the pressure
due to the line charge modulations at large distances as well.
Notice that the pressure for uniformly charged lines (\ref{P0d}) 
goes down more slowly. 
\end{itemize}

We conclude that within the split of the total pressure onto its particle
and deviation parts (\ref{contributions}), implied by the contact value
theorem, both short-distance and large-distance asymptotics due to
the surface charge modulations are dominated by the particle
term $\beta P_{\rm part}(d)$.
This term effectively diminishes the pressure in the absence of the
line charge modulation $\widetilde{P}_0(d)$ as was expected. 

\begin{figure}[tbp]
\centering
\includegraphics[clip,width=0.7\textwidth,angle=-90]{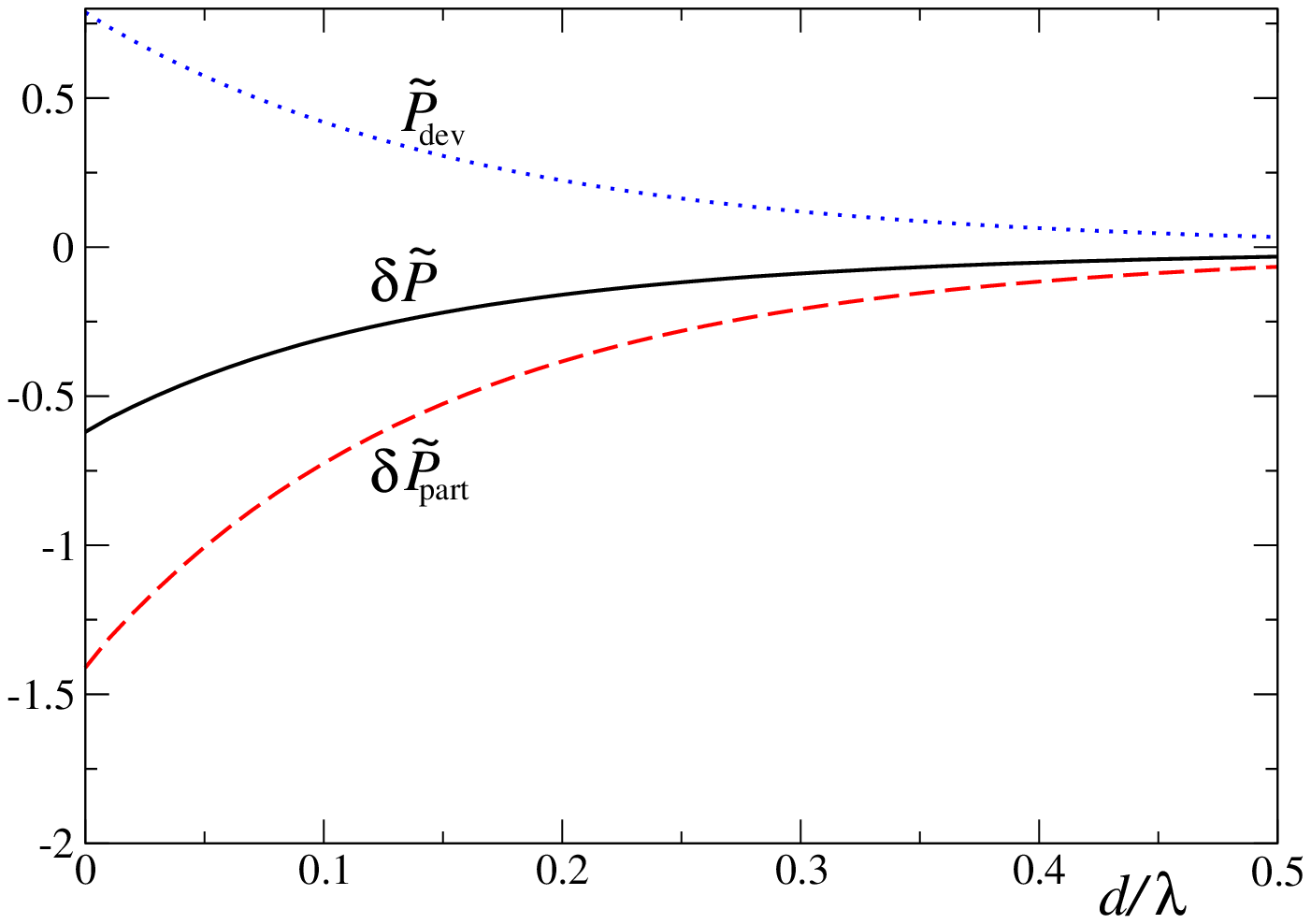}
\caption{The exactly solvable 2D model of two parallel EDLs
with counterions only, the symmetric line charge modulations
are given by (\ref{linechargemod}).    
The numerical results for various pressure contributions described
in the text, see equations (\ref{P1}), (\ref{P2}) and (\ref{P3}), as
the functions of the dimensionless interface distance $d/\lambda$.
The difference between the pressures with and without the line charge
modulations, $\delta\widetilde{P}(d)$, is always negative as it should be.}
\label{fig3}
\end{figure}

Keeping in mind that we consider the extreme value of the parameter
$\lambda \sigma = \frac{1}{2}$ preserving the exact solvability of
our model with the line charge modulation, let us consider also
the extreme value of the dimensionless amplitude of the
modulated surface charge density $\lambda A = \frac{1}{2}$
constrained by (\ref{extreme}).
This means that the line charge densities
\begin{equation} \label{linechargemod}
\lambda \sigma_L(y) = \lambda \sigma_R(y) = \frac{1}{2}
\left[ 1 + \cos\left( \frac{2\pi}{\lambda}y \right) \right]
\end{equation}  
are positive, except for the points $y=\pm \lambda/2, \pm 3\lambda/2,\ldots$
where they vanish.
The numerical results for various pressure contributions as the functions
of the dimensionless interface distance $d/\lambda$ are presented
in figure \ref{fig3}.
In particular,
\begin{equation} \label{P1}
\delta\widetilde{P}_{\rm part}(d) \equiv
\widetilde{P}_{\rm part}(d) - \widetilde{P}_0(d)
\end{equation}
(dashed curve) is the (always negative) difference between the particle part
of the pressure and the pressure with the uniform line charge densities
($\lambda A = 0$),
\begin{equation} \label{P2}
\widetilde{P}_{\rm dev}(d) = \frac{\pi}{4} {\rm e}^{-2\pi d/\lambda}
\end{equation}  
(dotted curve) is the (always positive) deviation part of the pressure and
\begin{equation} \label{P3}
\delta\widetilde{P}(d) \equiv \widetilde{P}(d) - \widetilde{P}_0(d)
= \delta\widetilde{P}_{\rm part}(d) + \widetilde{P}_{\rm dev}(d)
\end{equation}
(solid curve) is the difference between the pressures with and without
the line charge modulations.
It is seen that $\delta\widetilde{P}$ is negative in the displayed interval
of the distances $d/\lambda$ as was expected.

\renewcommand{\theequation}{5.\arabic{equation}}
\setcounter{equation}{0}

\section{Conclusion} \label{Sec5}
Rigorous or exact results in the thermodynamics of Coulomb fluids,
valid for any spatial dimension, are rare.
The contact value theorem belongs to such results.
In its most general formulation, it relates the pressure to
the averaged one- and two-body densities inside the particle domain.
In this paper, we generalized the contact value theorem to EDLs with
modulated surface charge densities, see equation (\ref{final3}) for
the geometry of one EDL and equations (\ref{contributions})-(\ref{final6})
for two parallel EDLs at distance $d$.
These equations involve the particle density at the wall interfaces and
the charge density of particles inside the whole domain $\Lambda$,
in formal analogy with the jellium systems characterized by the volume
background charge density \cite{Choquard80,Totsuji81}.
In the case of two parallel EDLs the total pressure is obtained as
the sum of the particle and surface-charge deviation parts, see equations
(\ref{contributions})-(\ref{final6}).
We tested these two contributions on a symmetric 2D model with
counterions only which was solved exactly in \cite{Samaj22}.
It turns out that the (negative) particle part dominates over the
(positive) deviation part which explains the diminution of the total
pressure due to the surface charge modulation.

As concerns a potential application of the present contact value theorem
in future, it may motivate someone to establish a general proof about
the pressure diminution due to the surface charge modulation.
A step towards the general proof might be the consideration of small
surface charge deviations $\delta\sigma_L({\bf y})$ and
$\delta\sigma_R({\bf y})$ on the left and right interfaces, respectively.
In this limit, the contributions of the surface charge deviations to the
particle number and charge densities can be treated perturbatively
around the system with the uniform surface charge densities within
the linear response theory.
A technical problem is that the first nonzero contribution to the pressure,
bilinear in the deviations $\delta\sigma_L({\bf y})$ and
$\delta\sigma_R({\bf y})$, involves three-body densities of the
unperturbed system which are difficult to deal with.

Another possible extension of the present work
is to consider dielectric discontinuities between the walls
and the medium in which charged particles move and Casimir-like forces.
This requires the inclusion of the interaction of particles with their
image charges (one-wall geometry) or an infinite array of image charges
(two-wall geometry) which ultimately leads to the presence of
higher-order particle densities in the formalism.

\ack
The support received from VEGA Grant No. 2/0089/24 and Project
EXSES APVV-20-0150 is acknowledged.

\end{document}